\documentclass[11pt,a4paper]{article}
\usepackage{graphicx}
\usepackage{epstopdf, epsfig}
\usepackage[applemac]{inputenc}
\usepackage[T1]{fontenc}
\usepackage{amssymb}
\usepackage{amsmath,amsfonts}
\usepackage[francais, english]{babel}
\usepackage{color}
\usepackage{subfigure}
\usepackage{authblk}

\title{Elastohydrodynamic wake and wave resistance}

\begin{document}

\author[1]{Maxence Arutkin}
\author[1,$^{\dagger}$]{Ren\'e Ledesma-Alonso}
\author[1,2]{Thomas Salez}
\author[1]{\'Elie Rapha\"el}

 \affil[1]{Laboratoire de Physico-Chimie Th\'eorique, UMR CNRS 7083 Gulliver, ESPCI Paris, PSL Research University,
10 rue Vauquelin, 75005 Paris, France} 
\affil[2]{Global Station for Soft Matter, Global Institution for Collaborative Research and Education, Hokkaido University, Sapporo, Hokkaido 060-0808, Japan.} 
\affil[$^{\dagger}$]{e-mail: rene.ledesma-alonso@espci.fr} 

\date{\today}
\maketitle

\begin{abstract}
The dynamics of a thin elastic sheet lubricated by a narrow layer of liquid is relevant to various situations and length scales. In the continuity of our previous work on viscous wakes (\cite{rene}), we study theoretically the effects of an external pressure disturbance moving at constant speed along the surface of a thin lubricated elastic sheet. In the comoving frame, the imposed pressure field creates a stationary deformation of the free interface that spatially vanishes in the far-field region. The shape of the wake and the way it decays depend on the speed and size of the external disturbance, as well as the rheological properties of both the elastic and liquid layers. The wave resistance, namely the force that has to be externally furnished in order to maintain the wake, is analyzed in detail.
\end{abstract}

Interfacial phenomena lead to qualitatively different behaviours from those encountered in bulk materials. In fluid mechanics and soft matter, this includes in particular the existence of surface waves. As an example, water waves have fascinated a large number of physicists and mathematicians for many decades. Among them, Lagrange derived the equation of water waves (\cite{darrigol}), and Kelvin described the wake behind a ship (\cite{kelvin}) -- characterized by the universal angle of $19.7\, \ensuremath{ ^\circ}$. This observation continues to trigger fundamental questions (\cite{rabaud,darmon}). Moreover, in the context of atomic-force microscopy and thin viscous films, the surface wake might directly be used as a new kind of nanorheological probe (\cite{Alleborn2007,nano,rene}). It may as well play a crucial role in biolocomotion, as demonstrated by the case of water striders that propel themselves using surface waves (\cite{bush}). For all the phenomena introduced above, and in fact many more~(\cite{demery}), the disturbance creates waves and thus radiates energy. As a consequence, the operator experiences a force opposing its motion, called the wave resistance (\cite{havelock,elie}). This aspect is crucial in the naval industry, through optimal design of boat shapes and recycling of the radiated energy for ecological purposes.

When a thin viscous film is coupled to an elastic layer, several interesting phenomena may happen. Classical hydrodynamic results have been revisited in this perspective, such as the capillary rise (\cite{duprat}), and the Saffman-Taylor viscous fingering (\cite{juel,stone,stone2}) for which the added compliance can prevent the instability. Exploring the physics of painting, the propagation of the peeling front in a plastic sheet atop a glycerine layer has been studied (\cite{hosoi}), as well as the flexible scraping of viscous fluids (\cite{clanet}). Besides, an emergent lift force exerted on a moving object near a brush (\cite{Sekimoto1993}), a soft (\cite{Skotheim2004,Snoeijer2013,salez}) or viscoelastic (\cite{pandey}) boundary was predicted and confirmed experimentally (\cite{saintyves}). Adhesive contact between a wet elastic sheet and a substrate also appears in a lot of physical and biological applications, and was shown to lead to patterns reminiscent of classical dewetting (\cite{mahadevan,carlson}). Measurements on small-scale systems using surface-force apparatus revealed striking substrate deformations (\cite{villey}), with obvious implications on the accuracy of nanorheological experiments. Finally, Brownian motion may be impacted as well by the inclusion of soft boundaries (\cite{daddi}).

The combination of both the wake and elastohydrodynamic physics introduced above naturally leads to a new class of interesting problems (\cite{blyth,guyenne}), with a broad range of applications in geophysics, biophysics, wave propagation, and engineering. For instance, seminal studies on elastohydrodynamic wakes were motivated by the waves generated by landing planes in Antarctica (\cite{parau,parau2}). We note that the inertia of the flow is a dominant ingredient in these works. 

In the present article, we study the displacement of an external pressure field above a thin elastic sheet covering a narrow viscous film. In the lubrication approximation, we compute the elastohydrodynamic waves and wake, as well as the wave resistance, in the continuity of our previous work on the viscocapillary case (\cite{rene}). An equivalent of the Bond number where elasticity replaces capillarity -- hereafter called the elastic Bond number -- appears to be a central dimensionless parameter of the problem. The elastohydrodynamic wake is plotted for a large range of speeds and elastic Bond numbers. Finally, in the low-speed and high-speed regimes, we provide analytical asymptotic results for the wave resistance.

\begin{figure}
\centering 
\includegraphics[scale=0.35]{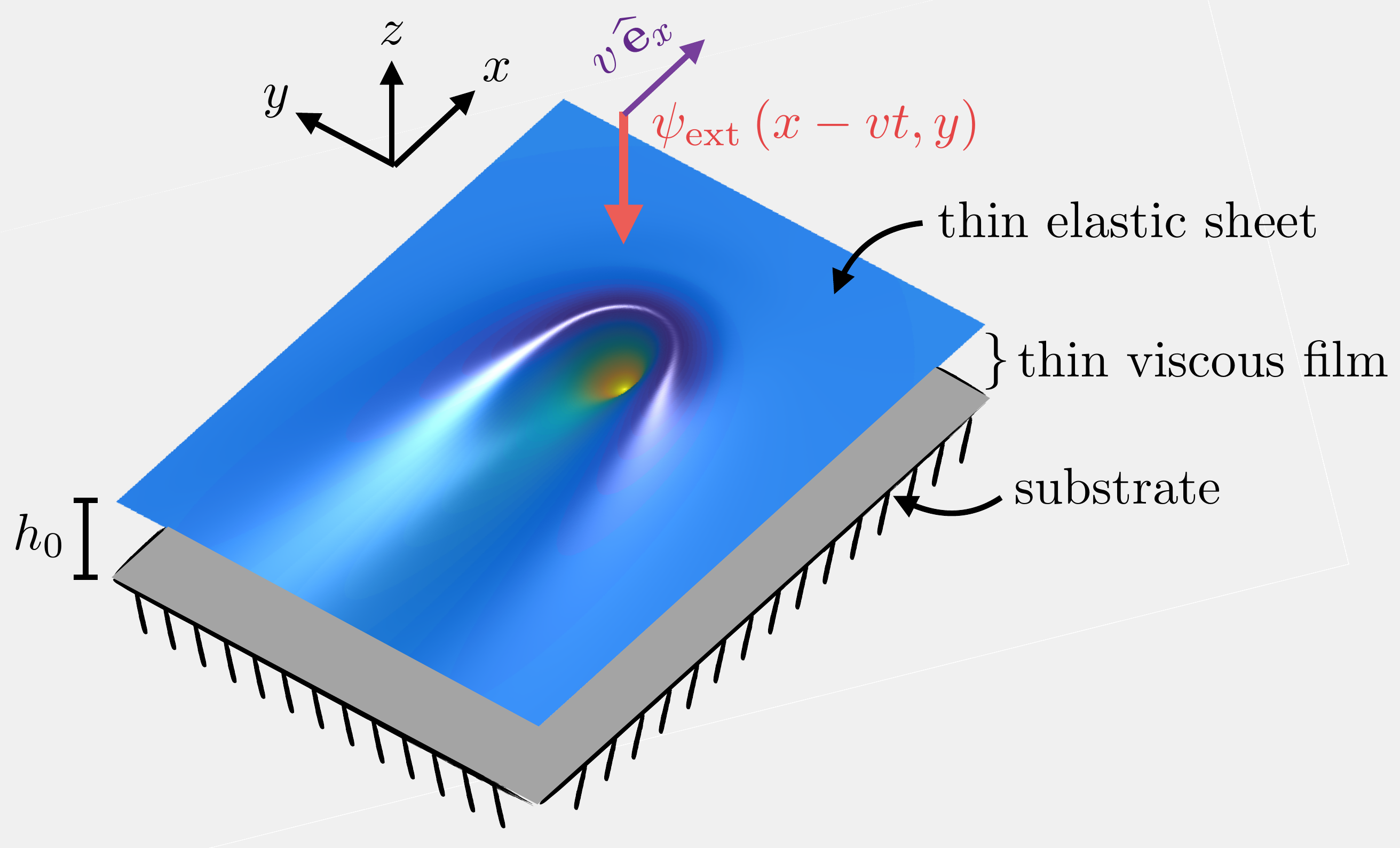}
\caption{Schematic of the elastohydrodynamic wake. The top elastic film reacts to an external disturbance $\psi_{\text{ext}}(x-vt,y)$ moving at constant speed $v\geq 0$ along the $x$ direction and time $t$.}
\label{schema}
\end{figure}

\section{Elastohydrodynamic lubrication model}
Let us consider a thin viscous film of thickness $h_{0}$ placed over a flat horizontal substrate, and covered by a thin elastic sheet of constant thickness $d\ll h_0$. As depicted in Fig.~\ref{schema}, an external pressure field $\psi_{\text{ext}}(x-vt,y)$, moving along the horizontal direction $x$ and time $t$ with a constant speed $v\geq 0$, is applied on the elastic sheet. A resulting non-constant profile $h\left(x,y,t\right)$ of the liquid-elastic interface, with respect to the substrate, is created. We note that the total two-layer profile $h+d$ has the same spatiotemporal variations as $h$, and we thus focus on the later only -- with no loss of generality. 

Invoking the incompressible Stokes equation and volume conservation, and considering no slip at both the substrate-liquid and the liquid-elastic interfaces, the following equation is yielded in the lubrication approximation (\cite{lister,stone,mahadevan,carlson}):
 \begin{eqnarray}
 \label{eqinit}
  \dfrac{\partial h}{\partial t}=\dfrac{1}{12\mu}\mathbf{\nabla}\cdot\left(  h^{3}\mathbf{\nabla} P_{\text{tot}} \right)\ ,
\end{eqnarray}
where $\mu$ is the dynamic viscosity, $\mathbf{\nabla}$ is the gradient operator in 2D Cartesian coordinates, and $P_{\text{tot}}(x,y,t)$ is the total pressure in the liquid. The latter is given by the addition of the bending stress, the hydrostatic pressure, and the external moving pressure field, and thus reads:
 \begin{eqnarray}
 \label{pinit}
P_{\text{tot}}=B\nabla^{4}h+\rho gh +\psi_{\text{ext}} \ ,
\end{eqnarray}
where $B=E d^{3}/[12(1-\nu^{2})]$ is the bending stiffness, $E$ and $\nu$ are respectively the Young's modulus and Poisson's ratio (\cite{landau}), $g$ is the acceleration of gravity, and $\rho$ is the density of the liquid. Note that we assumed the bending stresses to be dominant over the stretching ones. 

Taking $h_0$ and the gravito-elastic length $\kappa^{-1}_{\text{el}}=\left[B/\left(\rho g\right)\right]^{1/4}$ as the characteristic length scales in the vertical $z$-direction and in the $xy$ plane, respectively, the time $\tau =12 \mu/\left(\rho g \kappa^2_{\text{el}} h_{0}^{3}\right)$ as the characteristic time scale, and $P_{\text{ext}}=\kappa_{\text{el}}^{\,2}\iint dx\, dy \, \psi_{\text{ext}}$ as the characteristic pressure scale, we introduce the following dimensionless variables: $X=\kappa_{\text{el}} x$, $Y=\kappa_{\text{el}}y$, $H=h/h_{0}$, $T=t/\tau$, $\Psi=\psi_{\text{ext}}/P_{\text{ext}}$, and $\Gamma_{\text{el}}=P_{\text{ext}}/\left(\rho g h_0\right)$. In the limit of weak driving where $\Gamma_{\text{el}}\Psi\ll1$, and associated small deformation where $F=H-1\ll 1$, eqs.~(\ref{eqinit}) and~(\ref{pinit}) can be linearized and lead to the dimensionless elastohydrodynamic thin-film equation (\cite{flitton}) on the field $F(X,Y,T)$:
 \begin{eqnarray}
\dfrac{\partial F}{\partial T}=\Delta^{3}F+\Delta F + \Gamma_{\text{el}}\Delta\Psi\ ,
 \label{TFE6}
\end{eqnarray}
where $\Delta $ denotes the Laplacian operator in 2D Cartesian coordinates.

We restrict ourselves to stationary surface profiles in the comoving frame. Therefore, we introduce the new variable $U=X-V\, T$, and we define $F(X,Y,T)=\zeta\left(U,Y\right)$. In this context, eq.~(\ref{TFE6}) becomes:
\begin{equation}
\left[\dfrac{\partial^2}{\partial U^2}+\dfrac{\partial^2}{\partial Y^2}\right]^3\zeta+\left[\dfrac{\partial^2}{\partial U^2}+\dfrac{\partial^2}{\partial Y^2}\right]\zeta+V\dfrac{\partial}{\partial U}\zeta=-\Gamma_{\text{el}}\left[\dfrac{\partial^2}{\partial U^2}+\dfrac{\partial^2}{\partial Y^2}\right]\Psi \ .
\label{Eq:TFEHE}
\end{equation}
Finally, let us introduce the two relevant dimensionless parameters of the problem: the elastic Bond number $B_{\text{el}}=\left(a \, \kappa_{\text{el}}\right)^2$, where $a$ denotes the characteristic horizontal size of the external pressure field, 
and the reduced speed $V=\kappa_{\text{el}} \, \tau \, v$. We note that the vertical amplitude of the dimensionless disturbance field $\Gamma_{\text{el}}\Psi$ is a trivial dimensionless number in linear response theory, and we thus avoid discussing it further.

\section{Wake}
By definition, the wake is the solution $\zeta\left(U,Y\right)$ of eq.~\eqref{Eq:TFEHE}, for given disturbance field $\Gamma_{\text{el}}\Psi$ and reduced speed $V$. Invoking the two-dimensional Fourier transforms, $\widehat{\zeta}\left(K,Q\right)$ and $\widehat{\Psi}\left(K,Q\right)$, eq.~\eqref{Eq:TFEHE} becomes:
\begin{eqnarray}
\left[ iKV-(K^{2}+Q^{2})^{3}-(K^{2}+Q^{2}) \right]\widehat{\zeta}\left(K,Q\right)=\Gamma_{\text{el}}(K^{2}+Q^{2})\,\widehat{\Psi}\left(K,Q\right)\ .
\end{eqnarray}
Consequently, the solution reads:
\begin{equation}
\zeta\left(U,Y\right)=\dfrac{\Gamma_{\text{el}}}{4\pi^{2}}\iint\dfrac{(K^{2}+Q^{2})\exp\left[i(KU+QY)\right]\widehat{\Psi}\left(K,Q\right)}{iKV-(K^{2}+Q^{2})\left[1+(K^{2}+Q^{2})^{2}\right]} dKdQ \ .
\label{Eq:zeta}
\end{equation}

\begin{figure}
\centering
\includegraphics[scale=0.7]{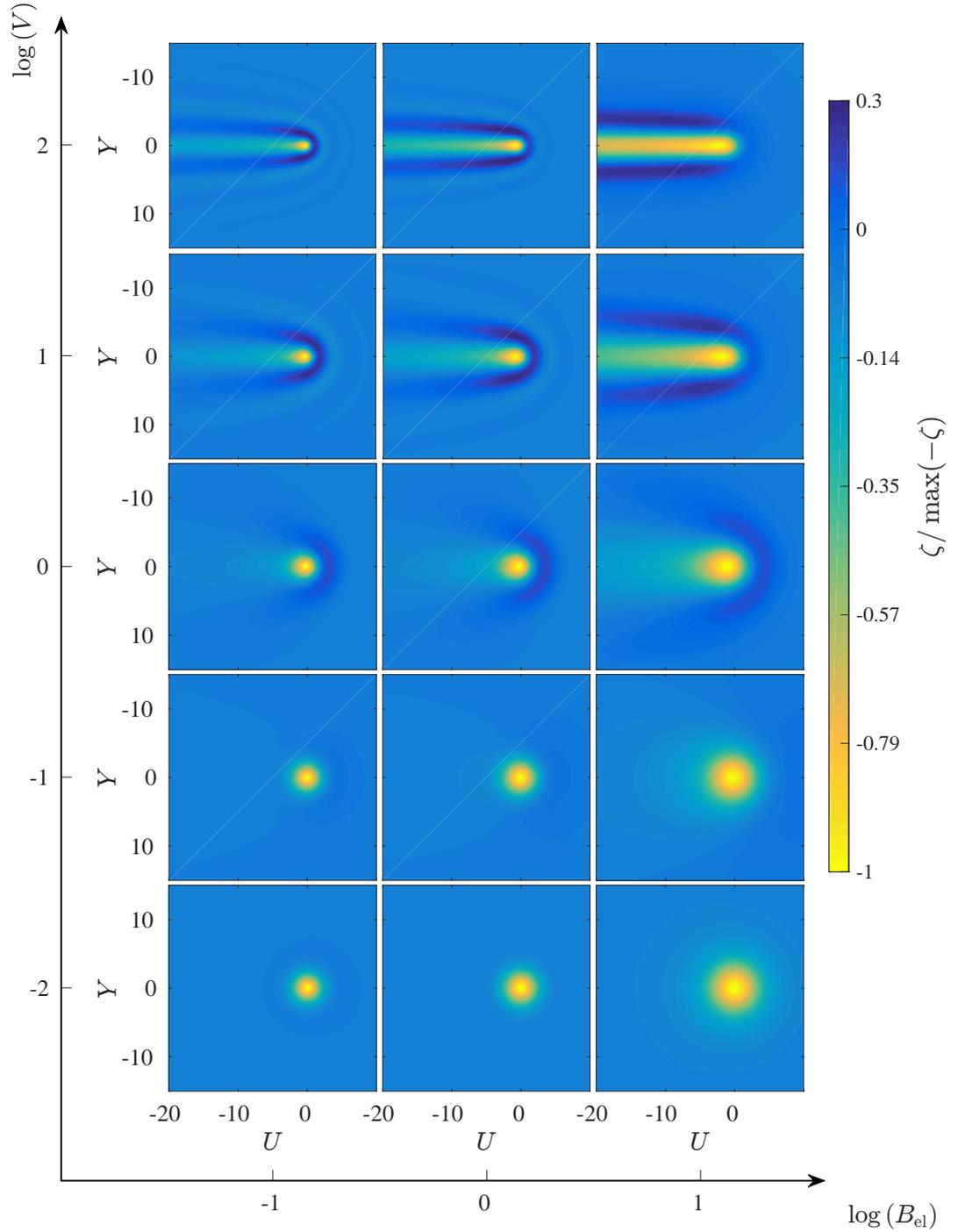}
\caption{Top view of the normalized wake created by a Lorentzian pressure field, computed from eqs.~\eqref{Eq:zeta} and \eqref{Eq:Lorentzian}, for different values of the elastic Bond number $B_{\text{el}}$ and the reduced speed $V$ as indicated. Since we use the comoving-frame variables, $U=X-V\,T$ and $Y$, the pressure field is centered at $U=0$ and the two-layer film travels at constant speed $V$ from right to left in each subfigure.} 
\label{wake}
\end{figure}

In order to fix ideas through a canonical example, we introduce the following axisymmetric Lorentzian pressure field and its Fourier transform:
\begin{subequations}
\begin{align}
\Psi\left(U,Y\right) &= \dfrac{\sqrt{B_{\text{el}}}}{2\pi\left(U^2+Y^2+B_{\text{el}}\right)^{3/2}} \ , \\
\widehat{\Psi}\left(K,Q\right) &= \exp\left[-\sqrt{B_{\text{el}}\left(K^2+Q^2\right)}\right] \ .
\end{align}
\label{Eq:Lorentzian}
\end{subequations}
The combination of eqs.~\eqref{Eq:zeta} and \eqref{Eq:Lorentzian} leads to the surface pattern at stake. A parametric study has been performed, sweeping a wide range of values for the reduced speed $V$ and elastic Bond number $B_{\text{el}}$, as summarized in Fig.~\ref{wake}. At low speed, the profiles are nearly symmetric and show a cavity below the the external pressure field. When the reduced speed reaches $V\sim1$, there is an accumulation of material at the front ($U>0$) and a stretching of the cavity behind the center of the pressure field ($U<0$). At high speed, comet-like shapes surrounded by undulations are observed, with an overall extent that can largely exceed the size of the pressure field.

\begin{figure}
\centering
\subfigure[]{
\centering
\includegraphics[width=0.45\textwidth,trim=58mm 70mm 60mm 30mm,clip]{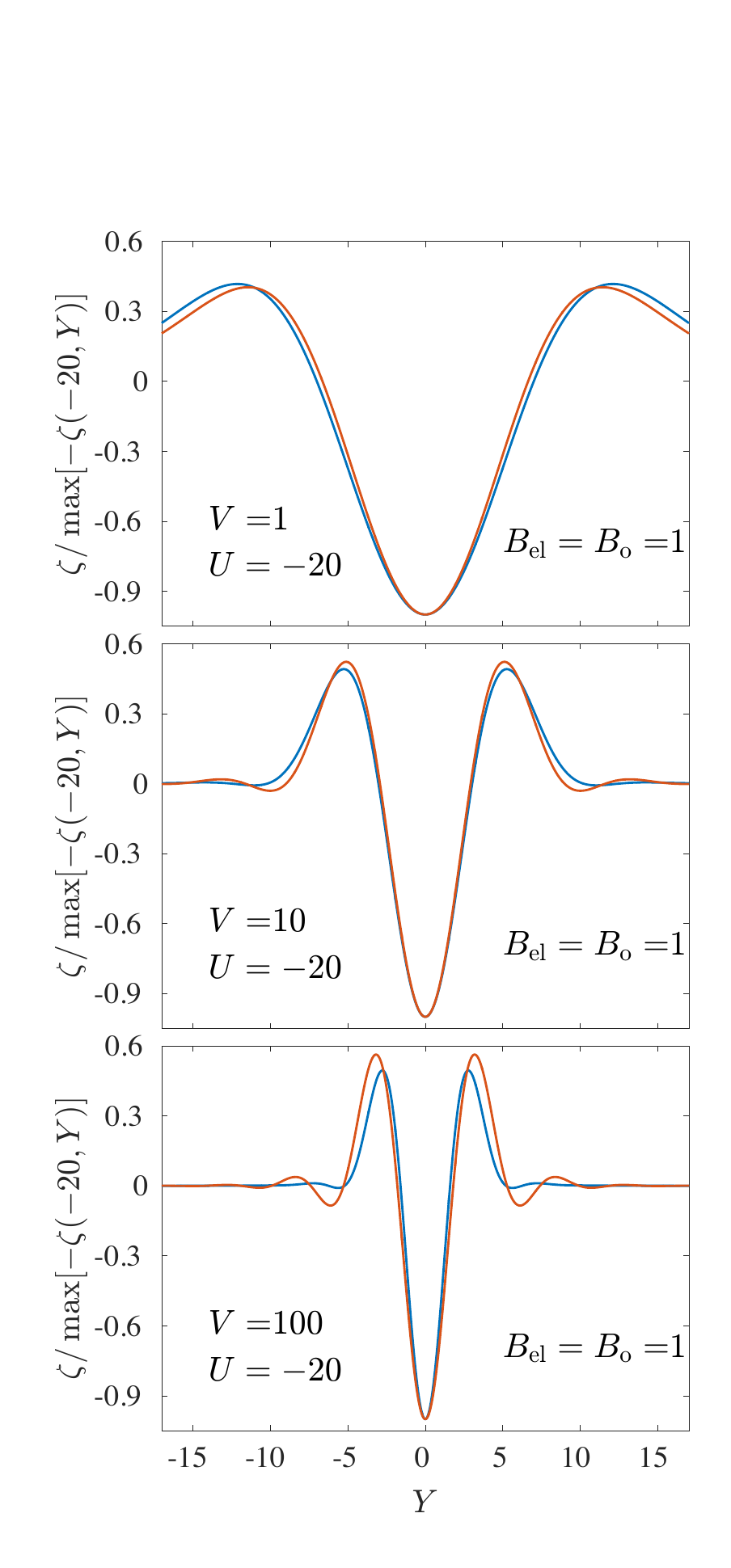}
\label{Fig:ZYComp}
}
\hspace*{5mm}
\subfigure[]{
\centering
\includegraphics[width=0.45\textwidth,trim=58mm 70mm 60mm 30mm,clip]{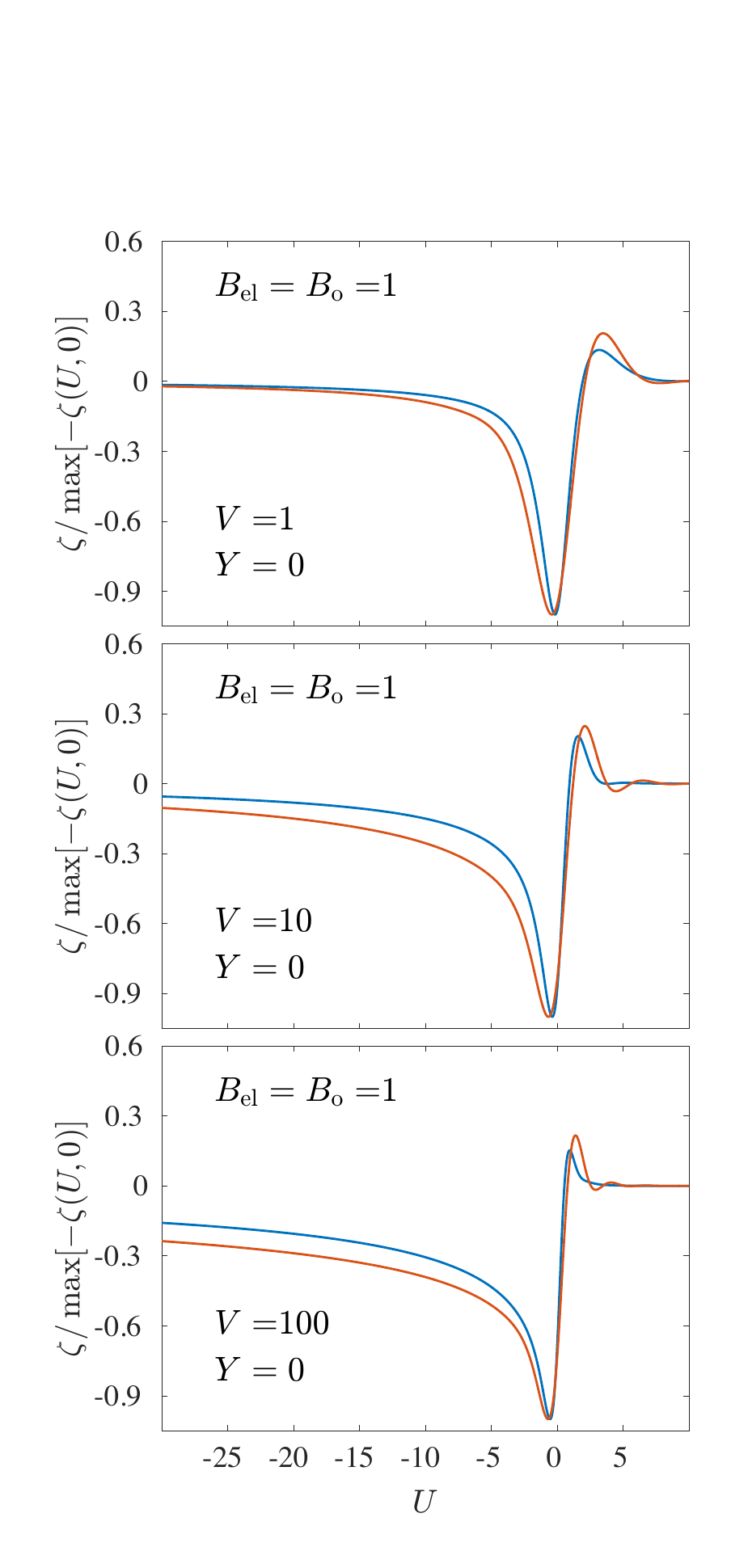}
\label{Fig:ZUComp}
}
\label{Fig:compTF46}
\caption{\subref{Fig:ZYComp} Normalized height profile as a function of the transverse coordinate $Y$, for a fixed longitudinal coordinate $U=-20$ and reduced speeds $V$ as indicated. \subref{Fig:ZUComp} Normalized height profile  as a function of the longitudinal coordinate $U=X-V\,T$, for a fixed transverse coordinate $Y=0$ and reduced speeds $V$ as indicated. In both panels, the orange curves correspond to the elastohydrodynamic case of eqs.~\eqref{Eq:zeta} and \eqref{Eq:Lorentzian} with $B_{\text{el}}=1$, whereas the blue curves correspond to the viscocapillary case (\cite{rene}) with a Lorentzian pressure distribution and $B_{\text{o}}=1$.}
\end{figure}

The comparison between the main features of the present elastohydrodynamic case and the ones previously reported for the viscocapillary case (\cite{rene}) is shown in Fig.~3. In the transverse cut of Fig.~\ref{Fig:ZYComp}, although the trends are similar, we observe that the oscillations are more pronounced in the elastohydrodynamic case -- consistent with the higher-order differential operator of the governing eq.~\eqref{Eq:TFEHE}. In the longitudinal cut of Fig.~\ref{Fig:ZUComp}, it is striking to notice the existence of oscillations at the front of the pressure disturbance in the elastohydrodynamic case, together with a longer decay length at the back. 

\section{Wave resistance}
As the pressure disturbance moves atop the elastic sheet, it generates the previously-discussed surface deformation which is intimately coupled to the motion of the underlying liquid. As a consequence of the viscous nature of the latter, a continuous dissipation of energy takes place, and the moving disturbance experiences a force opposing its motion. This so-called wave resistance $r$ is given by Havelock's formula (\cite{havelock,elie}):
\begin{eqnarray}
r=\iint \psi_{\text{ext}}\dfrac{\partial h}{\partial x} dx \, dy\ .
\end{eqnarray}
Naturally, the power $v r$ must be furnished by the operator in order to maintain a constant disturbance speed $v$ (\cite{rene}). 

With the notations introduced above, the dimensionless wave resistance $R$ reads:
 \begin{equation}
R=\dfrac{r\kappa_{\text{el}}}{\rho g h_0^2}=\Gamma_{\text{el}}\iint \Psi(U,Y)\dfrac{\partial \zeta}{\partial U}\, dU \, dY \ .
\label{Eq:WaveRes}
\end{equation}
Then, the substitution of eq.~\eqref{Eq:zeta} within eq.~\eqref{Eq:WaveRes} yields:
\begin{eqnarray}
R=\dfrac{\Gamma_{\text{el}}^{2}V}{4\pi^{2} }\iint\dfrac{K^{2}(K^{2}+Q^{2})\left\lvert\widehat{\Psi}\left(K,Q\right)\right\lvert^2}{K^{2}V^{2}+(K^{2}+Q^{2})^{2}\left[1+(K^{2}+Q^{2})^{2}\right]^{2}} \, dK\, dQ \ .
\end{eqnarray}
Invoking the polar coordinates, $K=\rho\cos\theta$ and $Q=\rho\sin\theta$, and assuming an axisymmetric pressure field $\widehat\Psi(\rho)$, allow us to integrate over $\theta$ and get the expression:
\begin{eqnarray}
R=\dfrac{\Gamma_{\text{el}}^{2}}{2\pi V }\int_{0}^{\infty}\left[1-\dfrac{\rho(\rho^{4}+1)}{\sqrt{V^{2}+\rho^{2}(\rho^{4}+1)^{2}}}\right]\left\lvert\widehat\Psi(\rho)\right\lvert^{2}\rho^{3} d\rho \ .
\label{eq:waveresist}
\end{eqnarray}

\begin{figure}
\centering
\includegraphics[scale=0.5]{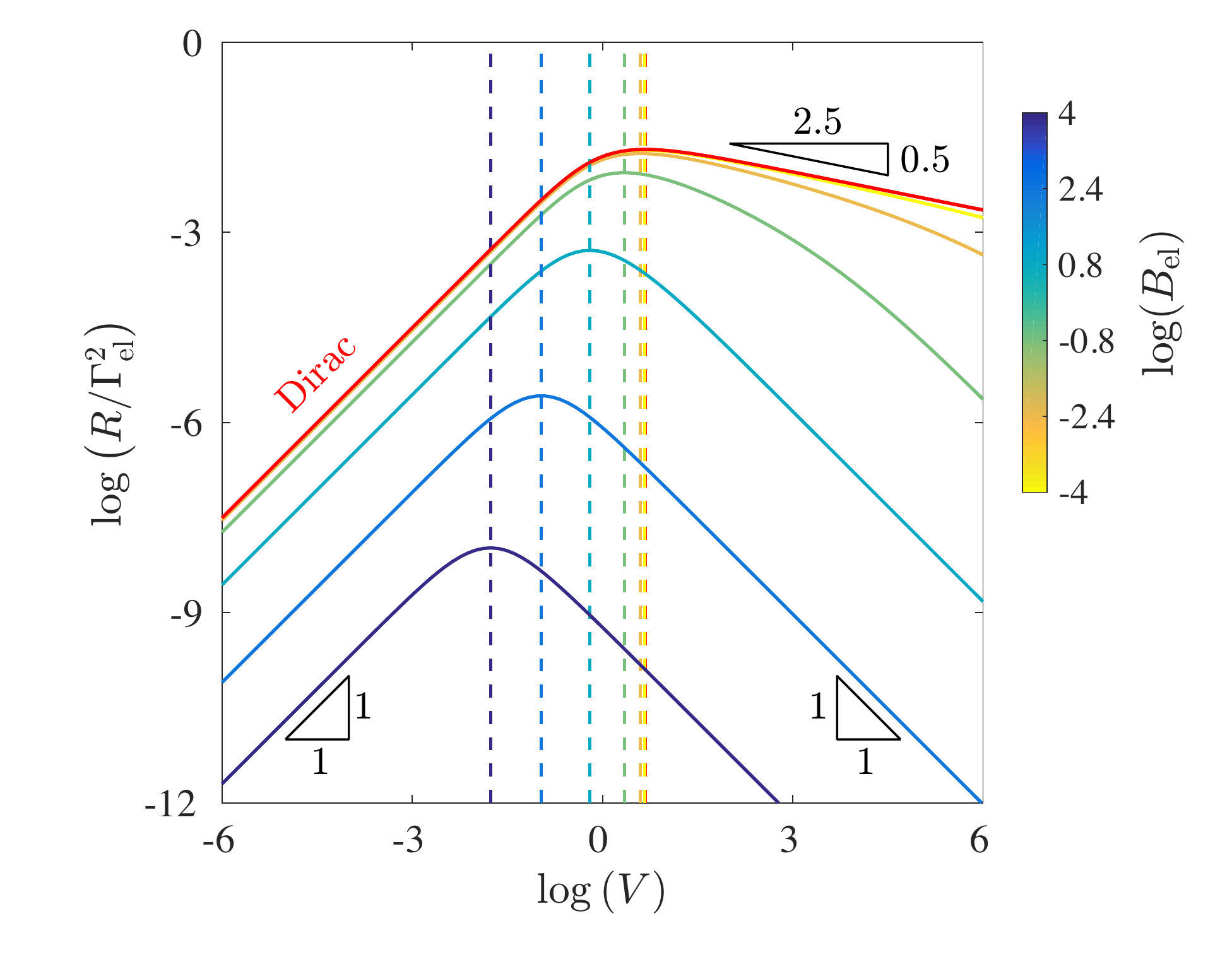}
\caption{Normalized wave resistance $R/\Gamma_{\text{el}}^{2}$, as given by eqs.~\eqref{Eq:Lorentzian} and~\eqref{eq:waveresist}, as a function of the reduced speed $V$, for various elastic Bond numbers $B_{\text{el}}$ as indicated. The Dirac limit ($B_{\text{el}}\rightarrow0$) is indicated. The vertical dashed lines indicate the position of the maximal wave resistance (see eq.~\eqref{vstar}), for each value of $B_{\text{el}}$.}
\label{Fig:Resistance}
\end{figure}

Equations~\eqref{Eq:Lorentzian} and~\eqref{eq:waveresist} allow us to compute numerically the wave resistance. The results are presented in Fig.~\ref{Fig:Resistance}. At low speed, the wave resistance shows a linear dependence with $V$; at high speed, it decreases inversely proportional to $V$; and at intermediate speed, it shows a maximum at the crossover between the two aforementioned regimes. In addition, the larger $B_{\text{el}}$, \textit{i.e.} the wider the pressure field, the lower the wave resistance; and the smaller $B_{\text{el}}$, \textit{i.e.} the narrower the Lorentzian pressure field, the more the wave resistance approaches the one associated with the Dirac pressure field ($\widehat{\Psi}=1$), as expected. Interestingly, the Dirac case shows a more gentle slope of $V^{-1/5}$ at high speed. In addition, we observe that the maximal wave resistance is bounded by the one of the Dirac case. Indeed, a Dirac pressure field excites all the wavelengths, whereas a Lorentzian pressure field cuts the large wavelengths off. 

\subsection{Low-speed regime}
As seen in the low-speed regime of Fig.~\ref{Fig:Resistance}, the wave resistance shows a linear rise with the speed $V$.
Indeed, using the Dirac pressure field, we get the following asymptotic expression from eq.~\eqref{eq:waveresist} as $V\rightarrow 0$:
\begin{eqnarray}
\label{dirc}
R\sim\dfrac{\Gamma_{\text{el}}^{2}V}{32 }  \ .
\end{eqnarray}
Similarly, using the Lorentzian pressure field of eq.~\eqref{Eq:Lorentzian}, the wave resistance reads in the low-speed regime:
\begin{eqnarray}
\label{rf}
R\sim\dfrac{\Gamma_{\text{el}}^{2}V}{4\pi }\mathcal F(B_{\text{el}}) \ ,
\end{eqnarray}
where:
\begin{equation}
\label{fbel}
\mathcal F(B_\text{el})=\int_0^{+\infty}\frac{\rho \exp\left(-2\sqrt{B_\text{el}}\rho\right)}{\left(\rho^4+1\right)^2}\, \text{d}\rho\ .
\end{equation}

In the limit of a wide pressure field, $B_{\text{el}}\gg 1$, one obtains:
\begin{eqnarray}
\mathcal F(B_{\text{el}})\sim\dfrac{1}{4B_{\text{el}}} \ .
\end{eqnarray}

\subsection{High-speed regime}
In the low-speed regime of Fig.~\ref{Fig:Resistance}, one has two distinct scaling behaviours with $V$.
Indeed, for the Dirac case, using the asymptotic development of eq.~\eqref{eq:waveresist} as $V\rightarrow \infty$, one gets:
\begin{eqnarray}
R\sim-\dfrac{\Gamma_{\text{el}}^{2}\Gamma(-2/5)\Gamma(9/10)}{20\pi^{3/2}}\dfrac{1}{V^{1/5}}  \ ,
\label{eq:WRHSdirac}
\end{eqnarray}
where $\Gamma(z)$ is the Gamma function, and has no relation with the dimensionless quantity $\Gamma_{\rm{el}}$, which has been defined in section 1.

In contrast, for the Lorentzian pressure field of eq.~\eqref{Eq:Lorentzian}, we obtain:
\begin{eqnarray}
R\sim\dfrac{3\Gamma_{\text{el}}^{2}}{16\pi B_{\text{el}}^{2}}\dfrac{1}{V} \ .
\label{eq:WRHSlorentz}
\end{eqnarray}
From the comparison of eqs.~\eqref{eq:WRHSdirac} and \eqref{eq:WRHSlorentz}, we remark that the wave resistance for a realistic finite-size pressure field decays much faster than for the singular Dirac pressure field.

\subsection{Maximal wave resistance} 
\begin{figure} 
\centering 
\includegraphics[scale=0.5]{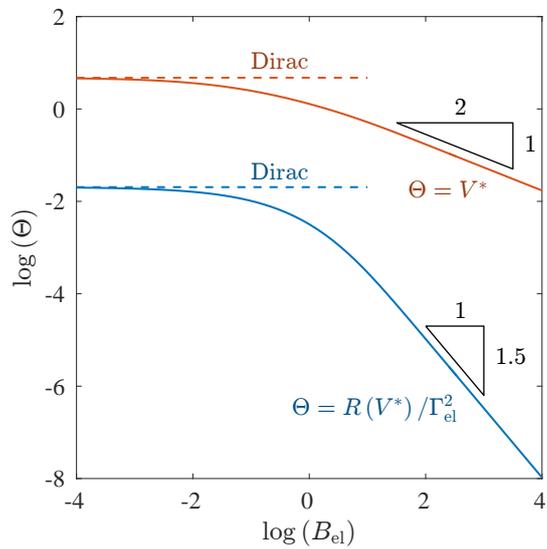} 
\caption{Computed values of the normalized maximal wave resistance $R(V^{*})/\Gamma_{\text{el}}^{2}$ (blue) and the corresponding speed $V^{*}$ (red), as functions of the elastic Bond number $B_{\text{el}}$, according to eqs.~\eqref{fbel} and \eqref{vstar}. The dashed lines represent the Dirac limits given in eqs.~\eqref{fin1} and~\eqref{fin2}.} 
\label{Fig:Max} 
\end{figure} 
As observed in Fig.~\ref{Fig:Resistance}, all the curves show a maximum for a certain speed $V^{*}(B_{\text{el}})$. An estimation of this speed can be obtained by balancing the low-speed and high-speed asymptotic expressions of the wave resistance in the Lorentzian case: 
\begin{eqnarray} 
\label{vstar} 
V^{*}\sim\sqrt{\dfrac{3}{4\mathcal{F}(B_{\text{el}})B_{\text{el}}^{2}}} \ . 
\end{eqnarray} 
Therefore, at high elastic Bond number, one finds: 
\begin{equation} 
\label{fin3} 
V^{*}\sim\sqrt{\dfrac{3}{B_{\text{el}}}} \ , 
\end{equation} 
and: 
\begin{equation} 
\label{fin4} 
R(V^{*})\sim\dfrac{\sqrt{3}\Gamma_\text{el}^2}{16\pi B_\text{el}^{3/2}} \ . 
\end{equation} 
The limit of the maximal wave resistance as $B_{\text{el}}\rightarrow0$ can be obtained by balancing the low-speed and high-speed asymptotic expressions of the wave resistance in the Dirac case: 
\begin{eqnarray} 
\label{fin1} 
V^{*}\sim\left[-\dfrac{8\Gamma(-2/5)\Gamma(9/10)}{5\pi^{3/2}}\right]^{5/6} \ , 
\end{eqnarray} 
and thus: 
\begin{eqnarray} 
\label{fin2} 
R(V^{*})\sim\dfrac{\Gamma_{\text{el}}^{2}}{32 }\left[-\dfrac{8\Gamma(-2/5)\Gamma(9/10)}{5\pi^{3/2}}\right]^{5/6} \ . 
\end{eqnarray} 
The asymptotic behaviors for low and high elastic Bond number, given by eqs.~\eqref{fin3} and \eqref{fin1} for $V^{*}$, and eqs.~\eqref{fin4} and \eqref{fin2} for $R(V^{*})$, are also represented in Figure.~\ref{Fig:Max}.

Surprisingly, this behaviour is qualitatively different from the viscocapillary case (\cite{rene}) for which the maximal wave resistance diverges in the Dirac limit. These results are summarized in Fig.~\ref{Fig:Max}. Therein, the maximal wave resistance decreases as $B_{\text{el}}^{-3/2}$ at large $B_{\text{el}}$, and saturates to a finite value in the Dirac limit. The corresponding speed $V^*$ decreases as $B_{\text{el}}^{-1/2}$ at large $B_{\text{el}}$, and saturates as well in in the Dirac limit.

\section{Conclusion}
We presented a theoretical investigation of the effects of a moving pressure disturbance above a thin elastic sheet placed atop a narrow viscous film. From the elastohydrodynamic lubrication model, we computed both the wake and the associated wave resistance experienced by the operator. A central dimensionless parameter of this study appeared to be the so-called elastic Bond number, measuring the ratio of gravity to elastic bending. As in the viscocapillary case, the wave resistance was observed to have a global maximum -- a point it might be interesting to control for energy-saving purposes. Finally, we conclude our study by an asymptotic analysis of the low-speed and high-speed regimes. These results are relevant to a wide class of geophysical, biological, and engineering problems, and may have implications in nanorheology, as well as wave propagation in metamaterials.

\section{Acknowledgments}
The authors thank Vincent D\'emery, Kari Dalnoki-Veress, Howard Stone and Antonin Eddi for interesting discussions. They acknowledge financial support from the Global Station for Soft Matter -- a project of Global Institution for Collaborative Research and Education at Hokkaido University.

\bibliographystyle{abbrv}
\bibliography{Arutkin2016bib}

\end{document}